# Parametrically squeezed states of microwave magnons in yttrium iron garnet films

M. Kostylev[1], A.B. Ustinov[2], A.V. Drozdovskii[2], B.A. Kalinikos[2], and E. Ivanov[1]

[1]*Department of Physics, The University of Western Australia, 6009 Crawley, Australia*
[2]*Department of Physical Electronics and Technology, Saint Petersburg Electrotechnical University "LETI", Saint Petersburg, 197376, Russia*

We demonstrate theoretically, and confirm experimentally, that nonlinear spin waves excited in thin yttrium iron garnet films are good candidates for squeezing vacuum quantum noise. The experimental demonstration is in the form of a measurement of spin-wave induced modulation instability (IMI) conducted in the classical regime. The experiment evidences strong phase locking of an idler wave parametrically generated in the film with a deterministic small-signal wave launched into the film from an external source. The theory predicts that the same behavior will be observed for vacuum quantum noise, resulting in squeezing of the noise.

Parametrically squeezed states of quantized fields [1,2] are of fundamental importance for the theory of quantum measurements [3,4] and have found applications in gravitational-wave detection and atom interferometers [1, 5] and in true random number generators [6]. They are also crucial for quantum computing, more precisely for continuous-variable quantum computing (CVC) [7, 8]. For instance, very recently, a proposition to employ CVC to build a quantum neural network has been put forward [9]. All these very important fields rely heavily on the quantum optics implementation of the Quantum Field Theory.

The demonstration of the parametric squeezing in optics dates back to the 1980ies [10, 11]. Both three- [10] and four- [11] photon parametric processes were employed to generate squeezed visible light. The four-wave parametric squeezing of the vacuum noise [11] is a quantum counterpart of a classical effect of the spontaneous modulation instability (SMI). SMI represents parametric amplification of noise in a nonlinear medium resulting in amplitude modulation of the pump wave. In fiber optics, SMI was observed in Ref. [12].

Coherent photon states can also couple parametrically to a quasi-classical pump wave. The same phase relationships are valid for the parametric amplification of the vacuum noise and the complex mode amplitude of a coherent state. The classical counterpart of the four-photon parametric amplification of the complex mode amplitude is the induced modulation instability (IMI). IMI represents parametric coupling of a classical deterministic signal to a pump wave. The coupling leads to an exponential increase in the signal amplitude with time and distance of the signal wave propagation. In fiber optics, IMI was observed in Ref. [13]. The same phase relationships remain valid for IMI, therefore observation of those relationships for classical waves may be considered as a prerequisite for a medium's ability to squeeze quantum noise. Due to the deterministic nature of all signals in the IMI case, technically, it is easier to observe phase relationships for IMI than for its "sister effect" – SMI.

The three- and four-wave parametric processes also take place in ferromagnets, but at microwave (MW) frequencies. They appear as interactions among quanta of spin waves (SW), which are called magnons. Nonlinear SW were widely studied in thin ferrimagnetic films of yttrium iron garnet (YIG) [16]. In particular, SMI of SW in YIG has been reported [15]. Importantly, SW in YIG films have much lower nonlinearity thresholds than light in the conventional optical fiber [14].

Parametric squeezing of magnons in ferro- and ferrimagnets has been discussed in the literature but only theoretically [17-19]. Ref. [17] considered three-magnon based squeezing, and [19] four-magnon one. Both works were carried out for an unconfined ferro-/ferrimagnetic medium. The works revealed that the squeezed states may exist in the medium for short periods of time. Contrary to refs. [17-19], we consider a different ferrimagnetic medium. These are the YIG films. A unique practical advantage of the film geometry is the easiness of excitation and detection of SW. Furthermore, the experimental observation of the spin-wave SMI was made using this particular medium. This suggests that from all possible geometries, the thin films may represent the best candidates for the experimental investigation of the squeezed magnon states in ferro-/ferrimagnetic materials. Accordingly, the aim of this paper is two-fold. Firstly, we carry out a theoretical analysis of the four-magnon processes in *YIG films* demonstrating that parametrically coupled magnon states are squeezed. Secondly, we carry out observation of the SW IMI in a YIG film and study phase dependences between the involved classical waves. The experiment confirms that thin YIG layers represent good candidates for future observation of the squeezed quantum states.

The goal of the theoretical analysis below is to derive expressions for the quadratures of quantum noise starting from a suitable classical model for the spin-wave IMI in YIG films. Previously, (see e.g., [10,14,16,20], and references therein), it was shown that four-wave self-modulation processes in YIG, such as the formation of SW envelope solitons, are well described by the nonlinear Schrödinger (NLS) equation model. In the case of SW, this equation reads:

$$i\left(\frac{\partial}{\partial t} + V_g \frac{\partial}{\partial z} + \gamma\right)c + \frac{D}{2}\frac{\partial^2 c}{\partial z^2} - T|c|^2 c = 0, \quad (1)$$

where $c$ is the scalar amplitude (envelope) of the wave, $V_g = \partial\omega(k)/\partial k$ is the wave's group velocity, $T$ is the nonlinear four-wave (self-modulation) coefficient, and $\gamma$ is the SW damping coefficient. $D = \partial^2\omega(k)/\partial k^2$ is the "dispersion coefficient". It represents the curvature of the SW dispersion curve $\omega(k)$, where $\omega$ is the SW frequency and $k$ is the SW wave number. The coefficients are calculated for a point $\omega(k_0)$ in the SW dispersion curve. The total SW field (envelope + MW carrier) reads $c(t,z) \exp(i\omega(k_0)t + i k_0 z)$.

Envelope solitons are formed in a medium, provided $DT<0$. Note that the same criterion is valid for the existence of SMI and IMI [21]. Furthermore, experimental investigations of SW solitons in YIG films are in good quantitative agreement with the model of Eq.(1) [22]. Therefore, we may start our theoretical analysis with Eq.(1).

The solution of the NLS equation, describing SMI/IMI, is: $c = c_0 + c_1 \exp(i\Omega t - i\kappa z) + c_2 \exp(-i\Omega t + i\kappa z)$. This Ansatz is a combination of an intense continuous-wave pump wave with an amplitude $c_0$ (and carrier frequency $\omega(k_0)$ and carrier wavenumber $k_0$), and two small-signal c.w. waves with amplitudes $c_1$ and $c_2$ ($|c_1|, |c_2| << |c_0|$) shifted in frequency by $-\Omega$ and $+\Omega$ respectively from the central frequency $\omega(k_0)$. The wave numbers of these two waves are shifted by $\pm\kappa$ from the carrier wave number $k_0$. Substituting the Ansatz into Eq.(1) and linearizing the result yields:

$$c_1(z) = \mu_1(z) c_1(z=0) + \nu_1 \overline{c}_2(z=0) ;$$
$$\overline{c}_2(z) = \overline{\mu}_2(z) \overline{c}_2(z=0) + \overline{\nu}_2 c_1(z=0), \quad (2)$$

where $z=0$ corresponds to the location where the three waves are launched into the film. This solution implies that the amplitudes of the small-signal waves scale linearly with the initial wave amplitudes $c_\alpha(z=0)$ (where $\alpha=1,2$). In the presence of magnetic losses in the film, there is no analytical solution to Eq.(1), and the coefficients $\mu_\alpha(z)$ and $\nu_\alpha(z)$ have to be found numerically. Note the presence of the non-vanishing cross-coupling coefficients $\nu_\alpha(z)$ in Eqs.(2). It evidences phase synchronism between $c_1$ and $c_2$.

The synchronism establishes as the two waves propagate along the film (i.e. in the $+z$ direction) and become parametrically amplified in the course of the propagation, due to their coupling to the pump wave. We now consider the beat signal of the two small-signal waves:

$$A = c_1 \exp\{i[\omega(k_0)+\Omega]t - i[k_0+\kappa]z\} +$$
$$+ c_2 \exp\{i[\omega(k_0)-\Omega]t - i[k_0-\kappa]z\} \quad (3)$$

We quantize it and Eqs.(2) and introduce its quadratures, $X = A\exp(i\phi) + A^\dagger \exp(-i\phi)$, $Y = i(A\exp(i\phi) - A^\dagger \exp(-i\phi))$ where $\varphi$ is some phase. Note that the loss term is present in Eq.(1). Therefore, our quantisation is analogous to solving a quantum Langevin equation for a medium with losses (Eq.(2) in [23]). We also assume a vacuum state $|0\rangle$ at the input of the film ($c_1(0)|0_1\rangle=0$, $c_2(0)|0_2\rangle=0$). This allows us to obtain fluctuations of the quadratures of the envelope signal at the frequency $\Omega$ [24]:

$$<0|X^2, Y^2|0> = |\mu_1|^2 + |\mu_2|^2 + |\nu_1|^2 + |\nu_2|^2 \pm$$
$$\pm 2|\nu_1||\mu_2|\cos(2\varphi + \arg(\nu_1) - \arg(\mu_2)) \pm \quad (4)$$
$$\pm 2|\mu_1||\nu_2|\cos(2\varphi + \arg(\mu_1) - \arg(\nu_2))$$

where the upper and lower signs correspond to the fluctuations of $X$ and $Y$ respectively. Here we introduced short-hand notations $\mu_i = \mu_i(z)$ and $\nu_i = \nu_i(z)$, and $\arg(...)$ denotes the phase angle of the respective complex-valued quantity.

The central result of this theory is shown in Fig. 1. This is an example calculation using Eq.(4), with $\mu_i$ and $\nu_i$ obtained from a numerical solution of Eq.(1) for a set of parameters used in the experiment described below: $D=-2\times10^{-3}$ cm$^2$/s, $T=1.5\times10^{10}$ s$^{-1}$, $V_g=4.4\times10^6$ cm/s, $\gamma=8.8\times10^6$ cm/s, and $\Omega/2\pi=25$ MHz. One sees that, theoretically, the quantum noise of one of the quadratures can be reduced to zero for some values of $\varphi$. Importantly, the fluctuations of the second quadrature are maximized for the same $\varphi$. This shows that the magnon quantum noise in a YIG film is squeezed. This is the main theoretical finding of this work. The squeezing is due to the process of four-magnon coupling of the three waves.

The strong periodic dependence of the quadrature fluctuations (Eq.(4) and Fig. 1) associated with the squeezing effect takes place because one quadrature of the input vacuum noise is de-amplified in some 180-degree wide range of the phase angle $\varphi$. This leads to suppression of the respective output noise quadrature. For the remainder of the phase angles, fluctuations of this quadrature are amplified. Furthermore, Eqs.(2) have the same form for both creation-annihilation operators for magnons and classical SW amplitudes. All this implies that if a coherent quantum state, or a small-signal classical wave, is launched into a medium, the quadratures of the output signal will be functions of $\varphi$ with a shape very similar to one from Fig. 1. Therefore, observation of a dependence of the same shape as in Fig. 1, for the quadratures of the output classical signal of IMI, must represent an experimental evidence that the parametrically amplified/de-amplified magnon quantum states are squeezed.

Quantum-optical experiments can be carried out at room temperature, because of the high energies of optical photons. This is not the case for MW magnons – in order to exclude the effect of thermal noise, experiments with YIG have to be carried at mK temperatures [25-28]. However, a study of IMI of classical SW can be carried out at room temperature.

To observe IMI, we use a MW circuit, seen in Fig. 2. Its main parts are a 5.7 μm thick YIG film grown with the liquid-phase epitaxy on top of a gadolinium gallium garnet (GGG) substrate and a MW balanced mixer. The distance of SW propagation in the film is 3.8 mm. The film is magnetized perpendicular to its plane by a field of 4338 Oe

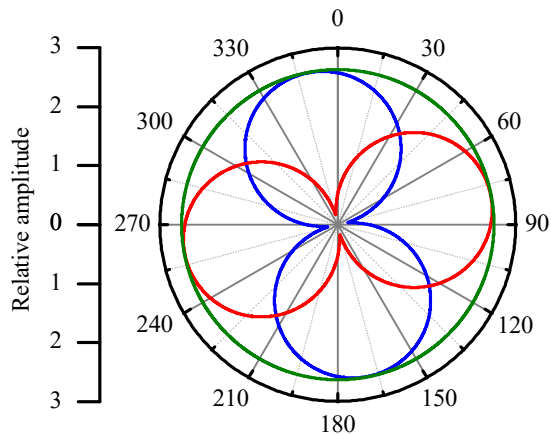

Fig. 1. Quadratures of the noise from the output of the film as a function of the phase angle $\varphi$. Red line: Expectation value for fluctuations of the $X$ quadrature ($\sqrt{<0|X^2|0>}$), blue line: one for the $Y$ quadrature ($\sqrt{<0|Y^2|0>}$). Green line: the sum of the expectation values ($\sqrt{<0|X^2+Y^2|0>}$).

A small-amplitude signal, of frequency 7915 MHz and power of +5 dBm, is applied to one input port of the power combiner (PC) from the MW generator SG2 (Fig. 2). The pump signal is fed into the second PC port from SG1. It has a frequency of 7889 MHz and power of +25 dBm. The generators are *phase locked*. Thus, at the film input ($z=0$), we have a combination of a weak signal wave ($S$), with $c_1$ and at $\omega_0+\Omega$=7915 MHz, and an intense pump wave ($P$), with $c_0$ and at $\omega_0$=7889 MHz ($c_0(z=0)\gg c_1(z=0)$).

A variable phase shifter controls the phase shift $\varphi$ between $P$ and the signal fed into the local oscillator (L) port of the mixer. The mixer-based receiver part of the circuit represents a MW analogue of the optical homodyne detector [11]. Therefore, it is able to register quadratures of the output signal of the film. The measurements show that the signal from the intermediate-frequency (I) port of the mixer has two frequency components – a DC one and one at $\Omega$ (26 MHz). We register the $\varphi$-dependence of the amplitude of the component at $\Omega$. This measurement is carried out with a MW spectrum analyzer. It yields the dependence shown in Fig. 3. The displayed quantity actually is one quadrature of the output signal of the film. The plot has the same shape as one from Fig. 1. This is the main experimental finding of this work.

The physics behind the shape of the plot from Fig. 3, is as follows. Four-magnon parametric coupling of $S$ to $P$ generates an idler wave ($I$) at $\omega_0-\Omega$=7863 MHz and with an amplitude $c_2$. Analyzing the frequency spectrum of the film output signal (not shown) with a MW spectrum analyzer confirms the presence of $I$. We expect $I$ to be phase locked to $S$. Because SG1 and SG2 (Fig. 2) are also phase locked, the phase locking of $S$ to $I$ results in deterministic beat of the two linear harmonic signals at the film output. The form of the beat signal is given by Eq.(3) for $z$=3.8 mm. The beat frequency is $\Omega$.

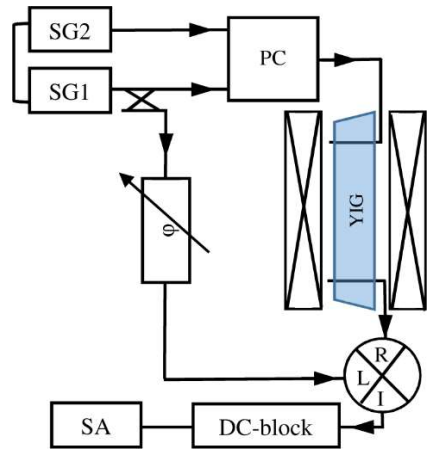

Fig. 2. Diagram of the experimental setup. SG 1 and SG 2 are MW generators generating signals at $\omega_0$ and $\omega_0+\Omega$ respectively, YIG is the YIG film, $\varphi$ is a variable phase shifter, L|R|I is a microwave balanced mixer ("L" denotes the local-oscillator port of the mixer and "I" is its intermediate-frequency port). SA is a Spectrum Analyzer, and PC is a power combiner.

The balanced mixer converts this signal into $c_1(z=0)[\mu_1(z=3.8\text{mm})\exp[i(\Omega t-(k_0+\kappa)z-\varphi)]+$ $\bar{v}_2(z=3.8\text{mm})\exp[-i(\Omega t+(k_0-\kappa)z-\varphi)]+c.c.]$. (Here we used Eqs.(2) and took into account that $c_2$=0 at $z$=0.) One sees that the signal from the I port of the mixer is a sine wave of frequency $\Omega$. In our experiment, we adjust the parametric gain (by carefully choosing the pump power and monitoring the film output spectrum with the spectrum analyzer) such that $|c_1(z=3.8\text{mm})|=|c_2(z=3.8\text{mm})|$. This implies that $|\mu_1(z=3.8\text{mm})|=|v_2(z=3.8\text{mm})|$. Accordingly, the $\Omega$ frequency output of the mixer is given by $2c_1(z=0)|\mu_1|\cos(\arg(\mu_1)/2-\arg(v_2)/2-\varphi)M(\varphi)$, where $M(\varphi)=\cos(\arg(\mu_1)/2-\arg(v_2)/2-\varphi)$, and all quantities are taken at $z=3.8\text{mm}$.

The form of $|M(\varphi)|$ is very similar to the $\varphi$-dependence from Eq.(4). (Note that Eq.(4) describes power of a signal and $|M(\varphi)|$ is a signal amplitude, therefore the dependence on $2\varphi$ in Eq.(4) and on $\varphi$ for $|M(\varphi)|$, but note the modulus sign in $|M(\varphi)|$.) In addition, the same quantities $\mu$ and $\nu$, responsible for quantum entanglement of the magnons at $\omega_0+\Omega$ and $\omega_0-\Omega$, enter $|M(\varphi)|$ and Eq.(4). Therefore, the quadratures of the IMI signal are characterized by the same two-fold symmetry, with respect to $\varphi$, as the vacuum noise quadratures. This is the sought evidence of the formation of a parametrically squeezed magnon state in the IMI process.

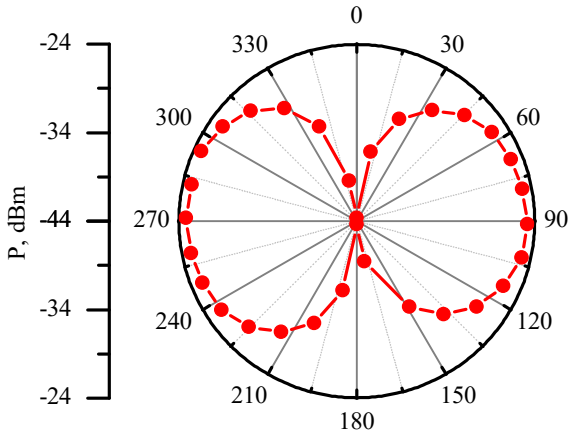

Fig. 3. Measured $\varphi$-angle dependence of the output of the balanced mixer at the frequency $\Omega$ (26 MHz).

In conclusion, we showed theoretically and confirmed experimentally that magnons in YIG films are good candidates for the generation of parametrically squeezed states at MW frequencies. The process of nonlinear four-particle interaction can be employed to this end. A classical signature of the medium's capability to form the squeezed states is IMI of the respective classical waves. We observed IMI of SW in a YIG film experimentally and measured quadratures of the beat of the signal and idler waves using an original experimental configuration. The configuration represents a MW analogue of the optical homodyne detector, widely used in quantum optics. This measurement delivered strong evidence that the quantum magnon states would be squeezed if the experiment were conducted in the single-magnon regime and at mK temperatures. The same experimental configuration, as in Fig. 2, may be employed in the future cryogenic experiment on measuring quadratures of the beat noise, with the only difference that SG2 will not be needed for the magnon noise measurement.

A Research Collaboration Award from the University of Western Australia and the Grant No. 14-12-01296-P from Russian Science Foundation are acknowledged.